\documentstyle[12pt]{article}

\newcommand{\mb}[1]{\mbox{\boldmath${\bf #1}$}}
\def\be{\begin{equation}}
\def\ee{\end{equation}}
\def\bea{\begin{eqnarray}}
\def\eea{\end{eqnarray}}

\begin{document}
\textheight 220mm
\textwidth 155mm
\topmargin -1cm

\noindent
\hspace*{10cm}  FTUV/96 -- 34\\
\hspace*{10cm}  IFIC/96 -- 42

\vskip 1cm

\begin{center}
{\bf SUPERSYMMETRIC QUANTIZATION OF YANG-MILLS THEORY AND POSSIBLE
APPLICATIONS} $\footnote{to appear in the Proceedings of the 2nd
Sakharov Conference on Physics, Moscow, May 20-25, 1996}$

\vskip 0.3cm
{F.G. SCHOLTZ  }

\vskip 0.3cm
{\em Department of Physics, University of Stellenbosch, 7600
Stellenbosch,\\ South Africa}

\vskip 0.3cm
{S.V. SHABANOV }

\vskip 0.3cm
{\em Department of Theoretical Physics, University of Valencia,
Burjassot, Moliner 50, E-46100, Spain and\\
Laboratory of Theoretical Physics, JINR, Dubna, Russia}
\end{center}

\begin{abstract}
We develop a new operator quantization scheme for gauge theories
where no gauge fixing for gauge fields is needed. The scheme
allows one to avoid the Gribov problem and construct a manifestly
Lorentz invariant path integral that can be used in the non-perturbative
domain. We discuss briefly an application of the method to Abelian
projections of QCD.
\end{abstract}

\section{Dynamics on the gauge orbit space}
Quantum dynamics of Yang-Mills fields is described by the Hamiltonian
\be
H_{YM} = \frac 12 \int d^3x \left( \mb{E}^2_a + \mb{B}^2_a\right)\ ,
\label{1}
\ee
where $\mb{E}_a$ and $\mb{B}_a$ are color electric and magnetic fields.
Physical wave functionals must be gauge invariant
$
\Psi_D[\mb{A}^U] = \Psi_D[\mb{A}]$,
$\mb{A}^U = U\mb{A}U^+ - iU\mb{\partial}U^+ $, $U^+U =UU^+=1,
$
which is ensured by the Dirac conditions (the Gauss law)
\be
\sigma^a_{YM} \Psi_D[\mb{A}] = \mb{D} ^{ab}\mb{E}_b \Psi[\mb{A}] = 0\ ,
\label{3}
\ee
with $\mb{D} ^{ab}$ being the covariant derivative in the adjoint
representation. The operator  $\sigma^a_{YM}$ of the Dirac constraint
is a generator of gauge transformations.

The Dirac states are functionals on the gauge orbit states
$[\mb{A}]_{ph}= [\mb{A}]/G$ where
$[\mb{A}]$ is the total configuration space and
$G$ is the gauge group, $G:\ \mb{A}\rightarrow \mb{A}^U$.
 In practical calculations, one needs to
introduce local coordinates on the orbit space, for instance, to go over
to the path integral formalism. In the conventional approach, one
fixes a gauge, say, $F(\mb{A}) = 0$, to parametrize the orbit space by
Cartesian "coordinates" satisfying the gauge condition. In this
approach it is assumed that the subspace $[\mb{A}]_{F=0}\subset
[\mb{A}]$ is isomorphic to the
orbit space. This assumption was shown \cite{gribov}
to be not justified because
the orbit space has a non-trivial topology, and therefore
there is no global coordinate system on it. In other words, one cannot
cover $[\mb{A}]_{ph}$ just by one Cartesian
coordinate patch $[\mb{A}]_{F=0}$
without singularities.

A natural resolution of the problem would be to use many patches that cover
$[\mb{A}]_{ph}$ without singularities. The latter is, however, inconvenient
from the practical point of view, especially in the path integral formalism.
The aim of the present paper is to develop an alternative approach where
dynamics on the orbit space is described without
gauge fixing for gauge fields.

\section{Supersymmetric quantization.}

Let us extend the initial configuration space $[\mb{A}]$ by adding to
the system two complex scalar (ghost)
fields: commutative $\phi$ and anticommutative (Grassmann) $\eta$
ones \cite{fr1}. We shall denote the set $(\phi, \eta)$ as $\Theta$.
 The ghost fields realize a unitary representation of the gauge
group
\be
\Theta \rightarrow \Theta^U = T_U \Theta\ ,\ \ \
\Theta^* \rightarrow \Theta^{*U} = \Theta^*T^+_U\ ,
\label{4}
\ee
where $T_U$ is the group element $U$ in the representation of the ghosts.
Consider quantum dynamics governed by the extended Hamiltonian \cite{fr2}
\bea
H = H_{YM} + H_{gh} = H_{YM} + [Q,R^+]_+
 = H_{YM}+ [Q^+,R]_+\ ,
 \label{5}\\
Q = i\int d^3x[(\eta^+,p_\phi )\! -\! (p_\eta^+,\phi)]\ ,
R = i\int d^3x[(p_\eta^+,p_\phi)\! +\! (\eta^+, M(\mb{A})\phi)];
\label{6}
\eea
here $(,)$ stands for an invariant scalar product in the
ghost representation space, $p_\Theta$ and $p^+_\Theta$ are
canonical momenta for $\Theta^+$ and $\Theta$, respectively,
satisfying the same statistics as the ghosts (i.e., the bosonic
ghost is quantized via a commutator, while the fermionic ghost
via an anticommutator),
and the linear, local and positive operator $M = M^+$
has the transformation
property
$
M(\mb{A}^U) = T_U M(\mb{A}) T^+_U$,
so that the extended Hamiltonian is gauge invariant.
The operators (\ref{6}) are nilpotent, $Q^2 = (Q^+)^2=0$,
$R^2 = (R^+)^2=0$. The ghost dynamics
exhibits an $N=2$ supersymmetry. From (\ref{5}), (\ref{6}) follows that
the ghost Hamiltonian $H_{gh}$, the odd operators
$Q, \ R$ and their adjoints form the $N=2$ superalgebra \cite{fr2}.
The operator $M$ can be chosen, for instance, as $-\mb{D}^2$.

Let $\sigma_{gh}^a$ be operators that generate the gauge transformations
(\ref{4}) of the ghosts. Gauge invariant states in the extended theory
are determined by
\be
\sigma^a \Psi_{ph}[\mb{A},\Theta,\Theta^*]=
(\sigma^a_{YM} +\sigma^a_{gh}) \Psi_{ph}[\mb{A},\Theta,\Theta^*] =0\ .
\label{9}
\ee
Our main observation is that \cite{fr2} the physical subspace
of the initial gauge theory determined by (\ref{3}) is isomorphic
to a subspace of the physical Hilbert space in the
extended theory that is selected by two supersymmetry conditions
\be
Q\Psi_{ph}[\mb{A},\Theta,\Theta^*]= Q^+\Psi_{ph}[\mb{A},\Theta,\Theta^*] =0\ ,
\label{10}
\ee
modulo states of zero norm.
Indeed, one can show that a generic solution to Eqs. (\ref{9}) and (\ref{10})
has the following form \cite{fr2}
\be
\Psi_{ph}[\mb{A},\Theta,\Theta^*] = \Psi_{gh}^{vac}[\mb{A},\Theta,\Theta^*]
\Psi_D[\mb{A}] + \Psi_0\ ,\ \ \
\Psi_0 = Q\Psi' = Q^+\Psi''\ ,
\label{11}
\ee
where $\Psi_{ph}^{vac}$ is the ghost vacuum determined by the equation
$
H_{gh}\Psi_{ph}^{vac} =0$.
Since $H_{gh}$ is a Hamiltonian of
an infinite dimensional supersymmetric oscillator,
the latter equation always has a normalized unique solution.
The norm of physical supersymmetric states is
equal to the norm of the Dirac states in the initial gauge theory
because i) $\Psi_0$ are orthogonal to the ghost vacuum (the ghost
vacuum is supersymmetric) and ii) $\Psi_0$ have zero norm (the
operators $Q$ and $Q^+$ are nilpotent).

For any operator of the form $O = O_{YM} + [Q, O'] = O_{YM} + [Q^+,O'']$,
where $O_{YM}$ is independent of the ghosts, one can prove the relation
\cite{fr2} $
\langle \Psi_{ph}|O|\Psi_{ph}'\rangle =$ $\langle
\Psi_{D}|O_{YM}|\Psi_{D}'\rangle$.
In particular,
\be
\langle \Psi_{ph}|e^{-itH}|\Psi_{ph}'\rangle =
\langle \Psi_{D}|e^{-itH_{YM}}|\Psi_{D}'\rangle\ ,
\label{14}
\ee
i.e., dynamics in the gauge invariant sector of the initial
theory is equivalent to supersymmetric gauge-invariant dynamics
of the extended theory.

\section{Avoiding the Gribov problem.}

Let $[\mb{A},\Theta,\Theta^*]$ be the total configuration space of
the extended theory. We have proved that the supersymmetric dynamics
on the extended orbit space $[\mb{A},\Theta,\Theta^*]_{ph} =
[\mb{A},\Theta,\Theta^*]/G$ is equivalent to dynamics on the
gauge orbit space $[\mb{A}]_{ph}$.
To resolve the aforementioned problem of parametrizing
 $[\mb{A}]_{ph}$, we introduce local coordinates on the
extended orbit space and consider supersymmetric dynamics in these
coordinates. The scalar bosonic ghost $\phi$
transforms homogeneously under the gauge transformations,
therefore, the space $[\mb{A},\Theta,\Theta^*]_{ph}$ can be parametrized
by means of imposing an algebraic gauge condition on it, 
$F(\phi)=0$, that is,
$[\mb{A},\Theta,\Theta^*]_{ph} \sim [\mb{A},\Theta,\Theta^*]_{F=0}$.
The supersymmetry transformations
generated by $Q$ and $Q^+$ become non-linear and involve $\mb{A}$
when projected on the hyperplane
$[\mb{A},\Theta,\Theta^*]_{F=0}$ \cite{fr3}.

Thus, the supersymmetric quantization scheme
allows one i) to avoid the Gribov ambiguity in the gauge field sector
(because there is no gauge condition imposed on gauge fields)
and ii) leads to the Lorentz invariant path integral (a
gauge condition is imposed on a Lorentz scalar $\phi$)
{\cite{fr1}, \cite{fr2}}.

On the canonical level the supersymmetric quantization has the further
advantage that the scalar product is well defined and no regularization is
required \cite{fr2}.  
This is in contrast to the conventional BRST quantization based on
gauge fixing\cite{hen}.

The main obstacle in the implementation of the supersymmetric quantization is
the difficulties presented by the renormalization of the theory.  This problem
has its origin in the fact that the supersymmetric quantization effectively
corresponds to the choice of a unitary gauge\cite{fr3}.

\section{Dynamical Abelian projection of gluodynamics.}

Recent numerical simulations of the string tension in the lattice
gluodynamics have shown that the main contribution to the Wilson loop
average comes from some specific configuratoins of gauge fields
\cite{suzuki}.
These configurations appear to be monopoles after partial gauge fixing
that restricts the gauge group to its maximal Abelian subgroup
\cite{thooft} and
give about 93 percents of the total string tension \cite{suzuki}.
This suggests
that gluodynamics is equivalent to some effective monopole dynamics
in the Abelian projection. To construct an effective
Lorentz invariant action of monopoles in the continuum gluodynamics,
one should, first, parametrize monopole configurations and, second,
resolve the Gribov problem arising after the Abelian projection, which
is difficult in the conventional approach.
In the framework of the supersymmetric quantization, one can
construct a new (dynamical)
Abelian projection \cite{s} where i) configurations relevant for
the path integral (monopoles) are classiafied in a gauge invariant way
and parametrized by a scalar (ghost) field in the adjoint representation,
and ii) their effective dymamics can be derived (because the Gribov problem
is no obstacle in this approach).

\end{document}